\documentclass[aps,prd,preprint,superscriptaddress,tightenlines,nofootinbib]{revtex4}

\usepackage{graphicx}
\usepackage{dcolumn}
\usepackage{bm}

\newcommand{\Ks}{K^0_S}
\newcommand{\Kl}{K^0_L}

\newcommand{\BF}{{\mathcal B}}
\newcommand{\Mbc}{M_{\rm BC}}

\newcommand{\Kz}{K^0}
\newcommand{\Kzbar}{\bar{K}^0}

\newcommand{\Amp}{{\mathcal A}}
\newcommand{\KpiCF}{D^0 \to K^- \pi^+}
\newcommand{\Kzbarpiz}{D^0 \to \Kzbar \pi^0}
\newcommand{\Kzbarpip}{D^+ \to \Kzbar \pi^+}
\newcommand{\Kzbarpi}{D \to \Kzbar \pi}
\newcommand{\KpiDCS}{D^0 \to K^+ \pi^-}
\newcommand{\Kzpiz}{D^0 \to \Kz \pi^0}
\newcommand{\Kpiz}{D^+ \to K^+ \pi^0}
\newcommand{\Kzpip}{D^+ \to \Kz \pi^+}
\newcommand{\Kzpi}{D \to \Kz \pi}

\begin{document}

\preprint{CLEO CONF 06-11}   

\title{Comparison of $D \to K^0_L \pi$ and $D \to K^0_S \pi$ Decay Rates}
\thanks{Submitted to the 33$^{\rm rd}$ International Conference on High Energy
Physics, July 26 - August 2, 2006, Moscow}

\author{Q.~He}
\author{J.~Insler}
\author{H.~Muramatsu}
\author{C.~S.~Park}
\author{E.~H.~Thorndike}
\author{F.~Yang}
\affiliation{University of Rochester, Rochester, New York 14627}
\author{T.~E.~Coan}
\author{Y.~S.~Gao}
\author{F.~Liu}
\affiliation{Southern Methodist University, Dallas, Texas 75275}
\author{M.~Artuso}
\author{S.~Blusk}
\author{J.~Butt}
\author{J.~Li}
\author{N.~Menaa}
\author{R.~Mountain}
\author{S.~Nisar}
\author{K.~Randrianarivony}
\author{R.~Redjimi}
\author{R.~Sia}
\author{T.~Skwarnicki}
\author{S.~Stone}
\author{J.~C.~Wang}
\author{K.~Zhang}
\affiliation{Syracuse University, Syracuse, New York 13244}
% ...
\collaboration{CLEO Collaboration} 
\noaffiliation

\date{July 24, 2006}

\begin{abstract} 
We present preliminary measurements of $D \to K^0_L \pi$ and $D \to K^0_S \pi$ 
branching fractions using 281 pb$^{-1}$ of $\Psi^"(3770)$ data at the CLEO-c experiment.  
We find that $\BF(D^0 \to \Ks\pi^0)$ is larger than $\BF(D^0 \to \Kl\pi^0)$, 
with an asymmetry of $R(D^0) = 0.122 \pm 0.024 \pm 0.030$.  For $\BF(D^+ \to \Ks\pi^+)$ 
and $\BF(D^+ \to \Kl\pi^+)$, we observe no measureable difference; the asymmetry is 
$R(D^+) = 0.030 \pm 0.023 \pm 0.025$.  Under reasonable theoretical assumptions, 
these measurements imply a value for the $D^0 \to K^\pm \pi^\mp$ strong phase 
that is consistent with zero.
The results presented in this document are preliminary.
\end{abstract}

\maketitle


\section{Introduction}
The $D^+$ meson, with quark composition $c\bar{d}$, decays by a Cabibbo-allowed decay
to $\bar{K^0}\pi^+$
($\bar{d}c \to \bar{d}s W^+_V \to (s\bar{d})(u\bar{d}) \Rightarrow \bar{K^0}\pi^+$), and
by a doubly-suppressed decay to $K^0 \pi^+$
($\bar{d}c \to \bar{d}d W^+_V \to \bar{d}du\bar{s} \to (d\bar{s})(u\bar{d}) \Rightarrow K^0\pi^+$).
Similarly, the $D^0$ meson, with quark composition $c\bar{u}$, decays by a Cabibbo-allowed
decay to $\bar{K^0}\pi^0$, and by a doubly-suppressed decay to $K^0\pi^0$. The
observable final states are {\it not} $\bar{K^0}$ or $K^0$, but rather $K^0_S$ or $K_L^0$.
As pointed out by Bigi and Yamamoto many years ago \cite{bigi}, interference between
Cabibbo-allowed and doubly-suppressed transitions leads to differences in the rates
for $D \to K_L^0 \pi$ and $D \to K_S^0 \pi$. Here we describe a search for differences
in the decay rates, both for $D^+\to K_S^0\pi^+$ vs. $D^+ \to K_L^0 \pi^+$ and for
$D^0 \to K_S^0 \pi^0$ vs. $D^0 \to K_L^0 \pi^0$. (Throughout, charge-conjugate modes
implied, except where noted.)

For these measurements we have used a sample of 281 pb$^{-1}$ $e^+e^- \to \Psi^"(3770)$
events, produced with the CESR-c storage ring and observed with the CLEO-c detector.

The data sample contains approximately 820,000 $D^+D^-$ events and 1,030,000 $D^0\bar{D^0}$ events,
as well as $e^+e^-\to u\bar{u},d\bar{d},s\bar{s}$ continuum events,
$e^+e^-\to\tau^+\tau^-$ events, Bhabha events, and $\mu$-pairs. The resonance $\Psi^"(3770)$
is below the threshold for $D\bar{D}\pi$, and so the events of interest, $e^+e^-\to
\Psi^"\to D\bar{D}$, have $D$ mesons with energy equal to the beam energy and a
unique momentum.

For the decays $D\to K_L^0 \pi$, we make no attempt to detect the $K_L^0$, as this is not feasible
with the CLEO-c detector.  Rather, we fully reconstruct a tag $\bar{D}$ on ``the other side,'' detect
the $\pi$, and compute the missing mass squared.  Our signal is a peak at the $K_L^0$ mass squared.

Explicitly, for $D^+ \to K_L^0 \pi^+$, we
reconstruct the $D^-$ in 6 decay modes.
We do this by requiring that the candidate $D^-$ has energy consistent with the beam
energy, and ``beam-constrained-mass''($\sqrt{E_{beam}^2-|\Sigma \vec{P_i}|^2}$) consistent
with the $D^-$ mass. Given a reconstructed $D^-$ meson, we require that the remainder of the event (the ``$D^+$ side'')
contain only one charged
track (for the $\pi^+$), and that any calorimeter clusters do {\it not} form
a $\pi^0$. We then compute the missing mass $M_X$ in the reaction
$e^+e^- \to D^- \pi^+ X$. The result is shown in Figure \ref{fig:KlpiMMsqFit}, where the $K_L^0$ peak
is evident.

For the decay $D^0 \to K_L^0 \pi^0$, we follow a similar procedure. We reconstruct
$\bar{D^0}$ in the decay modes $\bar{D^0} \to K^+ \pi^-$, $K^+ \pi^-\pi^0$, and
$K^+ \pi^- \pi^+ \pi^-$. Given a reconstructed $\bar{D^0}$ meson, we require that the
``$D^0$ side'' contains no charged tracks, only one $\pi^0 \to \gamma \gamma$, and no
$\eta \to \gamma \gamma$. We then compute the missing mass $M_X$ in the reaction
$e^+ e^- \to \bar{D^0} \pi^0 X$. The result is shown in Figure \ref{fig:mm2_1pi0_pull4_D}, with the $K_L^0$
peak evident.

For $D^+ \to K_S^0 \pi^+$, we use the result of an independent CLEO-c analysis that measures many $D^0$ and $D^+$ hadronic branching fractions, including $D^+ \to K_S^0 \pi^+$.~\cite{bib:Dhad}  This decay is directly reconstructed, using $K_S^0 \to \pi^+ \pi^-$.  Unfortunately, the analysis on the full 281 pb$^{-1}$ sample is not yet complete; the result we use is based on a 56 pb$^{-1}$ subset.  An updated result, with approximately half the uncertainty, will be available soon.

No previous CLEO-c analysis has measured $D^0 \to K_S^0 \pi^0$, so we have also analyzed this mode.  We directly reconstruct this mode, either tagged with a reconstructed $\bar{D}^0$ (``double tag") or not tagged (``single tag").

\section{Quantum Correlation in $D^0$ Decays}
The situation for $D^0 \to K_S^0 \pi^0$ vs. $K_L^0 \pi^0$ has an added feature.
When $D^0$ and $\bar{D^0}$ are pair-produced through a virtual photon($J^{PC}=1^{--}$),
they are in a quantum coherent state. Then the decays of $D^0$ and $\bar{D^0}$
will follow a set of selection rules. They cannot decay to CP eigenstates with the
same CP eigenvalue, if we ignore CP violation in $D^0/\bar{D^0}$ system. On the other
hand, decays to CP eigenstates with opposite CP eigenvalues are enhanced. Similarly,
all other final states are subject to such inteference effects. As a result, the
measured branching fractions in this $D^0/\bar{D^0}$ system differ from those of
isolated $D^0$ mesons. The measured branching fractions of the same mode by double
tag and single tag methods will also differ from each other, especially for CP
eigenstate modes.

The quantum correlation effects are shown in Table~\ref{table:QC}, where
`` f '' stands for flavored modes, `` X '' stands for everything, `` $S_+$ '' stands
for CP even modes, `` $S_-$ '' stands for CP odd modes, and

\begin{displaymath}
\centering
x \equiv \frac{m_1 - m_2}{\Gamma}
\hspace{.7in}
y \equiv \frac{\Gamma_1 - \Gamma_2}{2\Gamma}
\hspace{.7in} 
\frac{\langle f | \bar{D^0}\rangle}{\langle f | D^0 \rangle} = r_fe^{-i\delta}
\end{displaymath}

where $r_f$ is the amplitude ratio of ``wrong sign'' decay($D^0 \to K^+ \pi^-$ for example)
to ``right sign'' decay($D^0 \to K^- \pi^+$ for example) and
$\delta$ is the phase difference.

\begin{table}[hbp]
\centering
\caption{Single Tag and Double Tag yields for $C=-1$ $D^0 \bar{D^0}$ events, to
leading order in x and y.}
\label{table:QC}
\begin{tabular}{c|c|c}        \hline
  &$S_+$ & $S_-$\\ \hline
f &$NB_fB_{S_+}(1+r_f^2+2r_fcos\delta)$ &$NB_fB_{S-}(1+r_f^2-2r_fcos\delta)$ \\
X &$2NB_{S_+}(1-y)$              &$2NB_{S-}(1+y)$ \\  \hline
\end{tabular}
\end{table}

Since the CP eigenvalue of $K_S^0 \pi^0$ is odd and the CP eigenvalue of $K_L^0 \pi^0$ is even we can see from Table~\ref{table:QC} that we will overestimate the
$K_L^0 \pi^0$ branching fraction in the double tag method and underestimate the
$K_S^0 \pi^0$ branching fraction. For single tag measurements,
the effects are small since $y$ is tiny.

Our procedure is the following:
\begin{enumerate}
\item We measure the ``branching fraction'' for $D^0 \to K_S^0 \pi^0$, untagged. This
gives us $\mathcal B(D^0 \to K_S^0 \pi^0)(1+y)$. Since $y$ is very small,
$0.008\pm0.005$[PDG], we can correct for it, obtaining
$\mathcal B(D^0 \to K_S^0 \pi^0)$.
\item We measure the ``branching fraction'' for $D^0 \to K_S^0 \pi^0$, with three
different flavor tags. Each gives us
$\mathcal B(D^0 \to K_S^0 \pi^0)(1-2r_fcos\delta + r_f^2)$. Using
$\mathcal B(D^0 \to K_S^0 \pi^0)$ obtained from the untagged measuement, we obtain
$1-2r_fcos\delta + r_f^2$, for each flavor tag. Since $r_f^2$ is known for these three
flavor tags, from this we can compute $1+2r_fcos\delta + r_f^2$ for each flavor tag.
\item We measure the ``branching fraction'' for $D^0 \to K_L^0 \pi^0$, with the same
three flavor tags as those used for $D^0 \to K_S^0 \pi^0$. Each gives us
$\mathcal B(D^0 \to K_L^0 \pi^0)(1+2r_fcos\delta + r_f^2)$. Using the values of
$1+2r_fcos\delta + r_f^2$ for each tag obtained in (2), we obtain
$\mathcal B(D^0 \to K_L^0 \pi^0)$, for each of the three flavor tags.  These three measurements are then averaged for the final result.
\end{enumerate}

\section{$D \to \Ks\pi$ Measurements}
The value of $\BF(D^+ \to \Ks \pi^+)$ is taken from reference \cite{bib:Dhad}: (1.55 $\pm$ 0.05 $\pm$ 0.06)\%.

$\BF(D^0 \to \Ks \pi^0)$ is measured with two methods: single tag and double tag.

\subsection{Single Tag $D^0 \to \Ks \pi^0$}
Candidates for $D^0 \to K_S^0 \pi^0$ in untagged events were formed by combining
a $K_S^0$, reconstructed by a pair of charged tracks through the decay
$K_S^0 \to \pi^+ \pi^-$, and a $\pi^0$ from pairs of photons detected in the
CsI crystal calorimeter. The invariant mass of $K_S^0$ candidates was required
to be within 3 standard deviations of the known $K_S^0$ mass. The sideband of
$K_S^0$ mass is from 4 standard deviations to 7 standard deviations on both sides.
The invariant mass of $\pi^0$ candidates was required to be within 4 standard
deviations of the known $\pi^0$ mass. Photons with energy less than 30 MeV
were not considered. Both beam constrained mass and $\Delta E$ were required to be within
3 standard deviations of the nominal value. Two sideband subtractions were used to subtract the
background. $\Delta E$ sideband subtraction was used to subtract the continuum
and combinatoric background. $K_S^0$ mass sideband subtraction was used to subtract
the peaking background under $\Delta E$ and $M_{\rm BC}$ distributions. This peaking
background was formed from a real $D$ with the final state $\pi^+ \pi^- \pi^0$,
in which $m(\pi^+ \pi^-)$ happens to be within the $K_S^0$ mass window. This mode
is not decayed from a $K_S^0$ resonant state, so the mass of $\pi^+ \pi^-$ shows a
flat distribution.  The $K_S^0$ mass sideband subtraction removes this background
effectively.

All the yields of Monte Carlo and data in the signal region and the sideband region are
shown in Table~\ref{table:yields-ST}. By using the luminosity and cross section
of $e^+e^- \to D^0 \bar{D^0}$ , we get the number of $D$'s in data. Combining
with the efficiency from signal Monte Carlo, we get the branching fraction for
$D^0 \to K_S^0 \pi^0$ in the untagged data sample.

\begin{table}
\centering
\caption{Yields in both Monte Carlo (top) and data (bottom).
The Monte Carlo input branching fraction for
$D^0 \to K_S^0 \pi^0$ is $1.06\%$.}
\label{table:yields-ST}
\begin{tabular}{c|c}        \hline
Mode                &$D^0\to K^0_S \pi^0$\\ \hline
$Y_{signal region}$ & 128407\\
$\Delta E$ sideband & 6629.7\\
$K^0_S$  sideband   & 5312.3\\
Sideband subtracted & 116465\\
$N_{D}+N_{\bar{D}}$ & 36295180\\
Efficiency          & 28.94\%\\ \hline
$\mathcal B$        & 1.053$\pm$0.007\%\\\hline\hline

$Y_{signal region}$ & 8726\\   
$\Delta E$ sideband & 944.4\\
$K^0_S$  sideband   & 294.3\\
Sideband subtracted & 7487.2\\
$N_{D^0/\bar{D^0}}$ & 1013314\\\hline
$\mathcal B$        & 1.212$\pm$0.016\%\\\hline
\end{tabular}
\end{table}

\begin{table}
\centering
\caption{Systematic uncertainty for single tag $\mathcal B(D^0 \to K_S^0 \pi^0)$}
\label{table:SysError-ST}
\begin{tabular}{c|c}        \hline
$\Delta E$ cut        &0.5\% \\
Tracking efficiency   &0.7\% \\
$\Delta E$ sideband   &0.82\%\\
$K_S^0$ efficiency    &1.1\%\\
$K_S^0$ sideband      &0.28\%  \\
Cross section         &2.75\% \\\hline
$\pi^0$ efficiency    &\\\hline
All (without $\pi^0$ efficiency) &3.20 \\\hline
\end{tabular}
\end{table}

The systematic uncertainties are listed in Table~\ref{table:SysError-ST}.
The uncertainties due to $\pi^0$ reconstruction efficiency will cancel in the 
comparison of branching fractions for $D^0 \to K_L^0 \pi^0$, $D^0 \to K_S^0 \pi^0$,
we will not include that uncertainty here.
The dominant uncertainty comes from the cross section, which is used to calculate
the number of $D$'s. This cross section is base on the 56 pb$^{-1}$ dataset.  Soon, when
the 281 pb$^{-1}$ result comes out, this uncertainty will improve.

Combining all the results above, the single tag branching fraction for 
$D^0 \to K_S^0 \pi^0$, without $\pi^0$ systematic uncertainty, is 
$1.212\pm 0.016 \pm 0.039\%$.

\subsection{Double Tag $D^0 \to \Ks \pi^0$}
For the tagged branching fraction of $D^0 \to K_S^0 \pi^0$, $\bar{D^0}$
was fully reconstructed as $\bar{D^0} \to K^+ \pi^-$,
$\bar{D^0} \to K^+ \pi^- \pi^0$, or $\bar{D^0} \to K^+ \pi^- \pi^+ \pi^-$.
In the tagged sample, we reconstructed $D^0 \to K_S^0 \pi^0$ in the same way as
in the untagged case. All the requirements were unchanged, except there were
additional requirements regarding the tag side. The tag $\bar{D}^0$ was required to
be within 3 standard deviations in both the $\Delta E$ and $M_{\rm BC}$ distributions.
For tag mode $\bar{D^0} \to K^+ \pi^- \pi^0$, the energy of the tag-side $\pi^0$'s lower-energy shower
was required to be
greater than 70 MeV. This requirement made the background in $\Delta E$
distribution flatter and thus more suitable for the use of $\Delta E$ sideband
subtraction to get the number of $D$'s in the tag side.

In the tagged data sample, since the $K_S^0 \pi^0$ mode has relatively fewer neutral and
charged particles than an average $D^0$ decay, it is easier to reconstruct
the tag $\bar{D}^0$ when $D^0 \to K_S^0 \pi^0$, especially for the tags with more charged or neutral particles. Therefore, the branching fraction of the signal mode is biased in the subset of
the selected tag. By checking the Monte Carlo truth in the tagged sample, we obtained
a correction factor for this tag bias.

Tag side $\Delta E$ sideband subtraction was used to subtract fake $D$ events.
The signal side $K_S^0$ mass sideband was used to subtract peaking background under
signal side $\Delta E$ and $\Mbc$ distributions. Just as in the untagged case, the
peaking background was formed from a real tag $D$ with, on the signal side, the final state $\pi^+ \pi^- \pi^0$,
in which $m(\pi^+ \pi^-)$ happens to be within the $K_S^0$ mass window.

With the yields and effciencies from signal Monte Carlo, we computed the branching
fraction in Table~\ref{table:Ks-DT}.

\begin{table}[h]
\centering
\caption{Branching Fraction for $D^0 \to K_S^0 \pi^0$ in double tag method.
``s-s'' means sideband-subtracted for tags, and
background subtracted for signal yield.
``a'' is the correction factor for tag bias;  Error is
statistical only. The Monte Carlo input branching fraction for
$D^0 \to K_S^0 \pi^0$ is $1.06\%$.}
\label{table:Ks-DT}
\begin{tabular}{c|cc|cc|c|c|c}        \hline
&\multicolumn{2}{c|}{Tags} &\multicolumn{2}{c|}{Signal Yield}&MC &
Corrections \\
MC&raw &s-s &raw &s-s & eff(\%)& a & BR(\%)\\ \hline
$K \pi$
        &859919&855772 &3119&2958.5 &33.04 &1.00  &1.046$\pm$0.020\\
$K \pi \pi^0$
        &1186407&1147754&4815&4129.5&32.57 &1.014 &1.089$\pm$0.020\\
$ K 3\pi$
        &1268558&1228967&4480&4156.5&31.20 &1.033 &1.049$\pm$0.017\\\hline
Data&raw &s-s &raw &s-s & eff(\%)& a & BR(\%)\\ \hline
$K \pi$
        &48095 &47440  &172  &155 &33.04 &1.00  &0.989$\pm$0.088\\
$K \pi \pi^0$
        &67576 &63913  &248  &203 &32.57 &1.014 &0.975$\pm$0.082\\
$ K 3\pi$
        &75113 &71039.5&276  &256 &31.20 &1.033 &1.118$\pm$0.075\\\hline
\end{tabular}
\end{table}

The uncertainties due to data and Monte Carlo differences in  $\pi^0$ and $K_S^0$
reconstruction efficiencies are the same as in the untagged case. These two uncertainties
will cancel in the ratio of these two branching fractions. The uncertainties
due to $\Delta E$ sideband subtraction and $K_S^0$ sideband subtraction were also
estimated in a similar way as in untagged case.

The average of the results for the three tag modes is $(1.032\pm 0.047)\%$, 
significantly different from the untagged result, illustrating the effect of
quantum correlations. (Because the quantum correlation correction factor, 
$1-2rcos\delta+ r^2$, is tag-mode-dependent, this average is not otherwise
of interest.)

\section{$D \to \Kl\pi$ Measurements}
We measure the $D \to \Kl\pi$ branching fractions with a missing mass technique.  We reconstruct the tag $\bar{D}$ in 3 $\bar{D}^0$ modes and 6 $D^-$ modes, and we combine it with a $\pi^0$ or $\pi^+$ to form a missing mass squared.  The $D \to \Kl\pi$ signal is a peak at the $K^0$ mass squared ($($0.49772 $\rm GeV)^2$ = 0.24773 $\rm{GeV}^2$).

To remove $D \to \Ks\pi$ events, as well as other backgrounds, we require that the event contain no extra tracks or $\pi^0$'s beyond those used in the tag $\bar{D}$ and the $\pi$.  This veto removes about 90\% of $D \to \Ks\pi$ events and a few percent of $D \to \Kl\pi$ events.

\subsection{$D^+ \to \Kl\pi^+$}
We reconstruct tag $D^-$'s in the 6 decay modes $D^- \to K^+\pi^- \pi^-$, $K^+ \pi^- \pi^- \pi^0$, $\Ks \pi^-$, $\Ks \pi^- \pi^0$, $\Ks \pi^- \pi^- \pi^+$, and $K^+ K^- \pi^-$.  Candidates must pass $\Delta E$ and $\Mbc$ cuts.

The tag reconstruction efficiency is generally higher when the signal $D^+$ decays to $\Kl\pi^+$ than for generic $D^+$ decays because $\Kl\pi^+$ has only one charged particle and at most one calorimeter cluster.  This biases the sample of tagged events in favor of signal events.  Therefore, we include a factor in the branching fraction calculation to correct for this tag bias.  The factor, measured in Monte Carlo, is the ratio of the tag reconstruction efficiency when the $D^+$ decays to $\Kl\pi^+$ to the efficiency when it decays to anything else.

The efficiency for observing $D^+ \to \Kl \pi^+$, given that the tag was successfully reconstructed, is measured in signal Monte Carlo.  It is essentially the efficiency for finding the $\pi^+$.

\begin{figure}[tbp]
\begin{center}
 \includegraphics[width=1.0\textwidth]{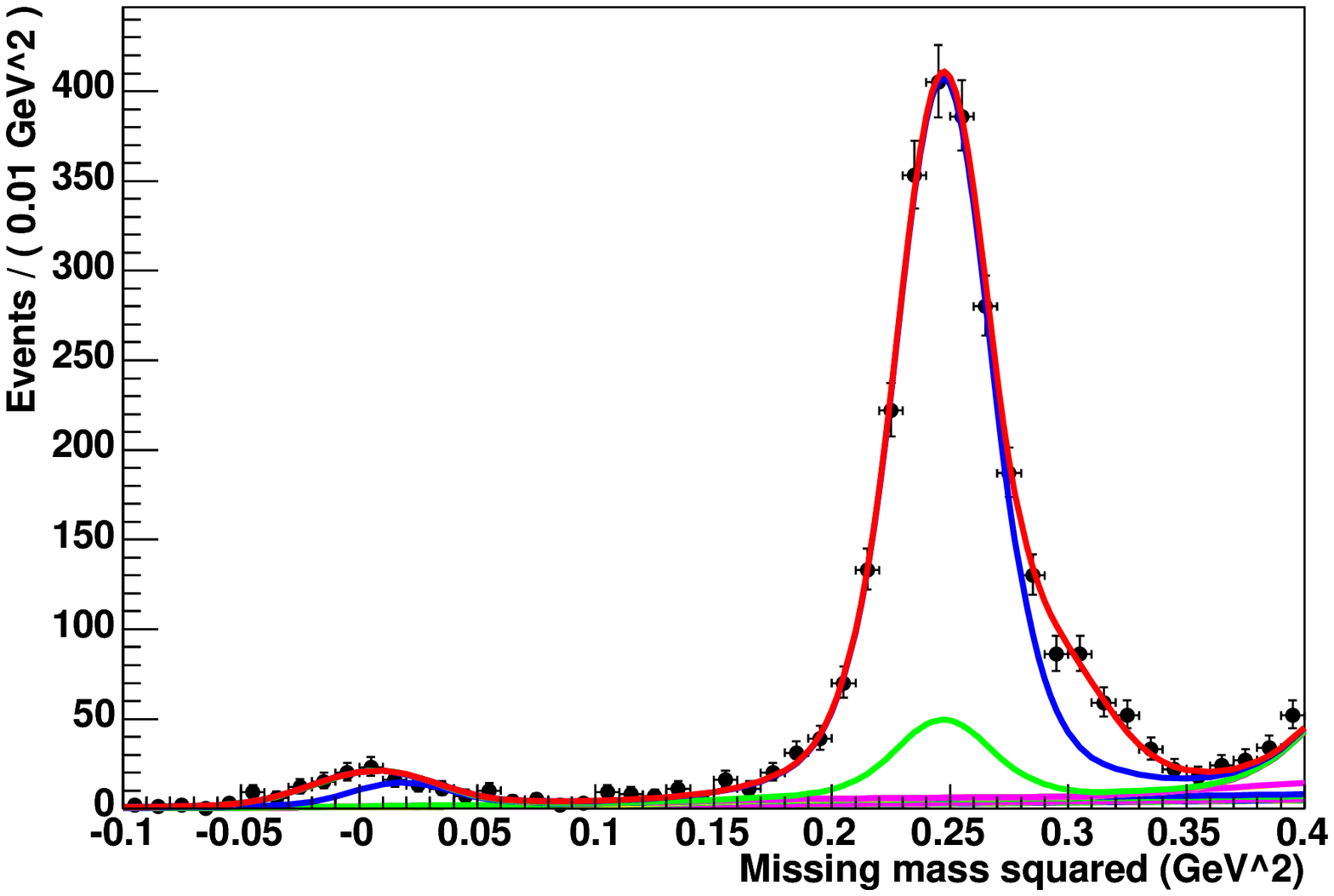}
\end{center}
\caption{Fit for $D^+ \to \Kl\pi^+$ yield using all tag modes.  The many colored lines represent the various background components, added cumulatively.  The green peak is the contribution of $D^+ \to \Ks\pi^+$ events that are not removed by the extra track and $\pi^0$ vetoes.}
\label{fig:KlpiMMsqFit}
\end{figure}

The missing mass squared distribution, with all tag modes added together, is shown in Figure \ref{fig:KlpiMMsqFit}.  The lines show a fit to determine the signal yield; each line represents a background component added cumulatively.  The most prominent feature is the signal peak at the $K^0$ mass squared ($\sim$0.25 $\rm{GeV}^2$).  A number of backgrounds are also present.  First, fake $D^-$ candidates produce a background which is estimated from an $\Mbc$ sideband.  All of the other backgrounds come from other $D^+$ decays.  The largest of these are $D^+ \to \Ks\pi^+$ (green peak under the signal), $\eta\pi^+$ (peak on the right-side tail of the signal), $\pi^0\pi^+$ and $\mu^+ \nu_\mu$ (peak on the left of the plot), $\bar{K}^0 \pi^+ \pi^0$, and $\pi^+\pi^0\pi^0$.  The shapes and efficiencies of these backgrounds are determined from Monte Carlo, and their branching fractions are used with the efficiencies to determine the size of each.  Fortunately, the extra track and $\pi^0$ vetoes greatly reduce many backgrounds, such as $D^+ \to \pi^+ \pi^+ \pi^-$.  Overall, we find a signal yield of about 2000 events.

Although Figure \ref{fig:KlpiMMsqFit} shows all tag modes together, we actually fit each tag mode separately and calculate a branching fraction for each.  The 6 branching fractions are then averaged to produce the final result.  Table \ref{tab:KlpiYieldTable} shows the yields and efficiencies for each tag mode and the resulting branching fractions, without any systematic uncertainties or corrections.

\begin{table}[tbp]
\begin{center}
\begin{tabular}{|l|c|c|c|c|c|}
\hline
                                & Tag Efficiency      &                  &                  & $D^+ \to \Kl \pi^+$       & Branching        \\
Tag mode                        & Factor              & Efficiency (\%)  & $D$ yield        & yield                     & fraction (\%)    \\
\hline
$D^- \to K^+ \pi^- \pi^-$       & 0.9949 $\pm$ 0.0032 & 82.31 $\pm$ 0.19 &  80108 $\pm$ 342 &  967 $\pm$ 37             & 1.459$\pm$ 0.056 \\
\hline
$D^- \to K^+ \pi^- \pi^- \pi^0$ & 0.9579 $\pm$ 0.0055 & 81.55 $\pm$ 0.27 &  24391 $\pm$ 315 &  345 $\pm$ 22             & 1.662$\pm$ 0.108 \\
\hline
$D^- \to \Ks \pi^-$             & 0.9908 $\pm$ 0.0039 & 82.37 $\pm$ 0.21 &  11450 $\pm$ 144 &  132 $\pm$ 14             & 1.387$\pm$ 0.147 \\
\hline
$D^- \to \Ks \pi^- \pi^0$       & 0.9565 $\pm$ 0.0062 & 81.94 $\pm$ 0.37 &  25494 $\pm$ 404 &  323 $\pm$ 23             & 1.479$\pm$ 0.108 \\
\hline
$D^- \to \Ks \pi^- \pi^- \pi^+$ & 0.9552 $\pm$ 0.0050 & 81.33 $\pm$ 0.24 &  16739 $\pm$ 314 &  184 $\pm$ 16             & 1.291$\pm$ 0.114 \\
\hline
$D^- \to K^+ K^- \pi^-$         & 0.9870 $\pm$ 0.0038 & 81.09 $\pm$ 0.21 &   6892 $\pm$ 154 &   72 $\pm$ 11             & 1.271$\pm$ 0.195 \\
\hline
Sum or average                  &                     & 81.80 $\pm$ 0.09 & 165074 $\pm$ 723 & 2023 $\pm$ 54             & 1.456$\pm$ 0.040 \\
\hline
\end{tabular}
\end{center}
\caption{Results for $D^+ \to \Kl \pi^+$.  The branching fraction for each tag mode is calculated from the corresponding yields, efficiency, and tag efficiency factor.  This table does not include systematic uncertainties or corrections.}
\label{tab:KlpiYieldTable}
\end{table}

The systematic uncertainties are listed in Table \ref{tab:KlpiSystematics}.  A small correction is applied for the particle identification of the $\pi^+$ in $D^+ \to \Kl\pi^+$, in addition to the uncertainty.  The largest systematics arise from whether we allow the signal peak width to vary, from the extra track and extra $\pi^0$ vetoes, from the shape of the signal peaks, and from the statistical uncertainty and input branching fraction of the $D^+ \to \Ks\pi^+$ background.  The veto systematics include uncertainties on finding real tracks and $\pi^0$'s in background events, as well as on finding fake tracks and $\pi^0$'s in signal events.  The fake track and $\pi^0$ systematics are determined by looking for extra particles in fully-reconstructed $D\bar{D}$ events, in both data and Monte Carlo.

The branching fraction, with systematics, is ${\mathcal B}(D^+ \to \Kl  \pi^+)$ = (1.460 $\pm$ 0.040 $\pm$ 0.035 $\pm$ 0.009)\%.  The final uncertainty is the systematic uncertainty due to the input value of ${\mathcal B}(D^+ \to \Ks \pi^+)$.

\begin{table}
\begin{center}
\begin{tabular}{|l|c|}
\hline
Pion tracking                                       & $\pm$ 0.35\%      \\
Pion particle ID                                    & 0.30 $\pm$ 0.25\% \\
Tag reconstruction: signal vs. non-signal           & $\pm$ 0.2\%       \\
$D^+$ vs $D^-$ tags                                 & $\pm$ 0.5\%       \\
$\Ks$ veto systematics                              & $\pm$1.1\%             \\
Peak shapes                                 & $\pm$0.69\% \\
Fake $D^-$ background shape                 & $\pm$0.15\% \\
Fake $D^-$ background yield                 & $\pm$0.35\% \\
Background yields (except $\Ks\pi^+$)       & $\pm$0.49\% \\
$D^+ \to \Ks\pi^+$ efficiency \& statistics & $\pm$0.80\% \\
Fixed vs. floating peak width               & $\pm$1.63\% \\
Tail of signal peak                         & $\pm$0.25\% \\
\hline
\hline
Total                                       & $\pm$2.43\% \\
\hline
\hline
$D^+ \to \Ks\pi^+$ branching fraction       & $\pm$0.62\% \\
\hline
\end{tabular}
\end{center}
\caption{Systematics for $\BF(D^+ \to \Kl\pi^+)$. 
The ``total systematic'' does not include the $D^+ \to \Ks\pi^+$ branching fraction systematic.}
\label{tab:KlpiSystematics}
\end{table}

\subsection{$D^0 \to \Kl\pi^0$}
\begin{figure}
\centering
\includegraphics[width=14cm]{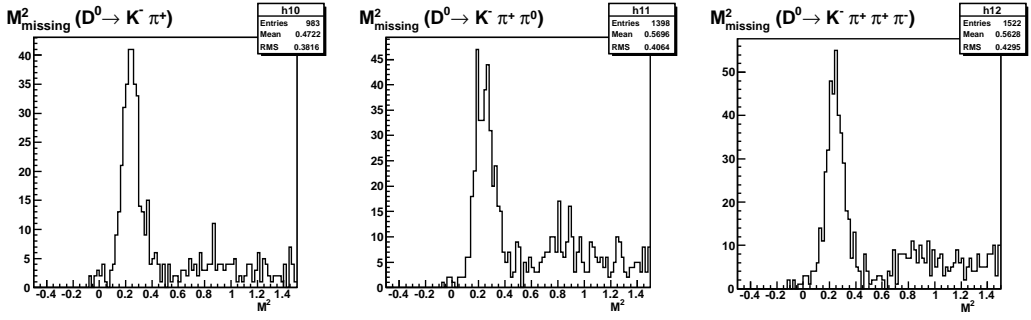}
\caption{Missing mass distribution in data after removing events with extra tracks, $\pi^0$'s, or $\eta$'s.
}
\label{fig:mm2_1pi0_pull4_D}
\end{figure}

For the tagged $D^0 \to K_L^0 \pi^0$ branching fraction measurement,
the same 3 $\bar{D^0}$ decay modes were selected with the same requirements as in the tagged
$D^0 \to K_S^0 \pi^0$ study. We require that there
are no tracks and only one desired $\pi^0$, and no $\eta$ on the other side. The invariant
mass of the $\pi^0$ was required to be within 4 standard deviations of the known $\pi^0$
mass (same as used before). The invariant mass of $\eta$ was required to be within 3
standard deviations of the known $\eta$ mass. After rejecting the events with any $\eta$ or track
or more than one $\pi^0$, we compute the missing
mass squared using the momentum of the $D$ and $\pi^0$, with both the $D$ and
$\pi^0$ masses constrained.
In Fig.~\ref{fig:mm2_1pi0_pull4_D}, we present the missing mass plots in data.

Since $D^0 \to K_L^0 \pi^0$ only has one observable neutral particle, $\pi^0$, the
branching fraction bias for $D^0 \to K_L^0 \pi^0$ is more apparent than that for
$D^0 \to K_S^0 \pi^0$. A correction factor was applied when computing the branching
fraction.

\begin{figure}[h]
\centering
\includegraphics[width=14cm]{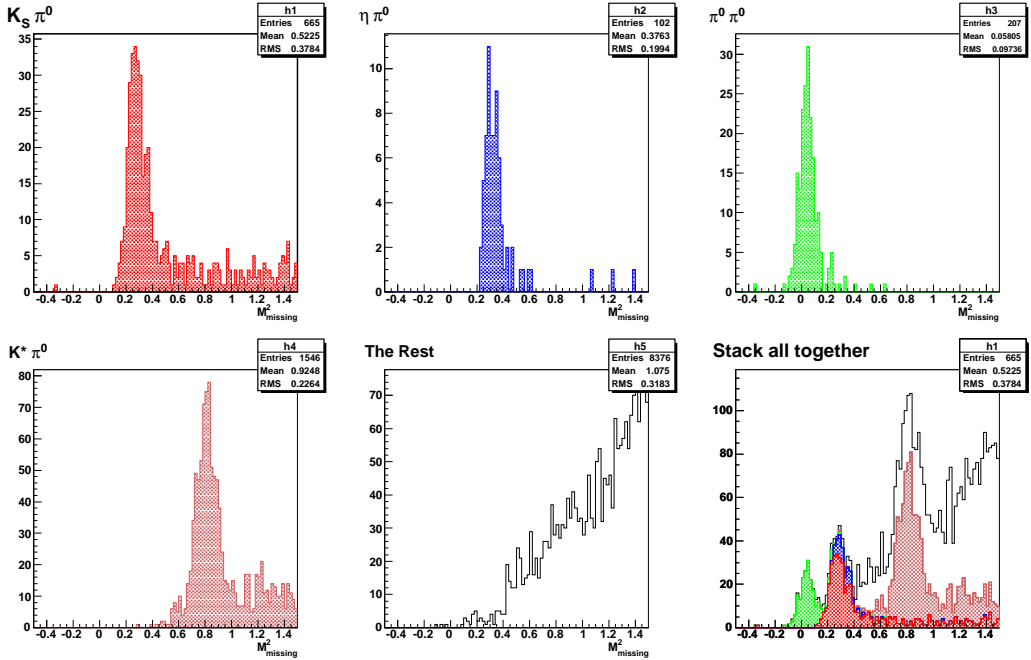}
\caption{Background for $D^0 \to K_L^0 \pi^0$. Upper left is $K^0_S \pi^0$,
upper middle is $\eta \pi^0$, upper right is $\pi^0 \pi^0$, lower left is $K^* \pi^0$,
lower middle is ``the rest'', and lower right is all cases stacked together.
}
\label{fig:mm2_bg2}
\end{figure}

A number of background channels appear in the missing mass squared plot --
$D^0\to K_S^0 \pi^0$, $D^0\to \eta \pi^0$, $D^0\to \pi^0 \pi^0$, $D^0\to K^{*}\pi^0$, and ``the rest" -- with
the $D^0\to \pi^0 \pi^0$ peak on the left side of our signal peak, and the $D^0\to K_S^0 \pi^0$
and $D^0\to \eta \pi^0$ peaks right under our signal peak, as shown in
Fig.~\ref{fig:mm2_bg2}. The total backgound is about 10\% in the signal region.

In order to get the signal and estimate the background, we define three regions
in $M_{missing}^2$: p-region ($-0.1\sim 0.1 \rm GeV^2 $), s-region ($0.1\sim 0.5 \rm GeV^2$),
and b-region ($0.8\sim 1.2 \rm GeV^2$).
The backgrounds were split into three groups:
$D^0 \to \pi^0 \pi^0$; $D^0 \to K_s^0 \pi^0$ and $D^0 \to \eta \pi^0$; and ``the
rest''. For $D^0 \to K_s^0 \pi^0$ and $D^0 \to \eta \pi^0$, we have no
experimental handles.  We just trust the Monte Carlo, and use it for the
subtraction. We use the yield in the p-region to estimate the
background from $D^0 \to \pi^0 \pi^0$, and the yield in the b-region to
estimate ``the rest.''

\begin{table}[h]
\centering
\caption{Branching fraction for $D^0 \to K_L^0 \pi^0$.
``s-s'' means sideband-subtracted for tags, and background-subtracted for signal yield.
``a'' is the correction factor for tag bias.  Error is
statistical only. The Monte Carlo input branching fraction for
$D^0 \to K_L^0 \pi^0$ is $1.06\%$.}
\label{table:all}
\begin{tabular}{c|cc|cc|c|c|c}        \hline
&\multicolumn{2}{c|}{Tags} &\multicolumn{2}{c|}{Signal Yield}&MC &
Corrections \\
MC&raw &s-s &raw &s-s & eff(\%)& a & BR(\%)\\ \hline
$K \pi$
       &791054 &787214 &5415 &4907.5&57.97 &1.00  &1.075$\pm$0.017\\
$K \pi \pi^0$
       &1102536&1066829&7378 &6554  &55.36 &1.037 &1.070$\pm$0.015\\
$ K 3\pi$
       &1171872&1136140&7390.5&6628 &52.37 &1.057 &1.054$\pm$0.014\\\hline
Data&raw &s-s &raw &s-s & eff(\%)& a & BR(\%)\\ \hline
$K \pi$
        &48095 &47440 &367  &334.8 &57.97 &1.00  &1.217$\pm$0.073\\
$K \pi \pi^0$
        &68000 &64280 &414.5&363.1 &55.36 &1.037 &0.984$\pm$0.058\\
$ K 3\pi$
        &75113 &71040 &466.5&418.0 &52.37 &1.057 &1.063$\pm$0.058\\\hline
\end{tabular}
\end{table}

After subtracting all the backgrounds, we get all the yields and computed branching
fractions in Table~\ref{table:all}.

Contributions to systematic uncertainty are categorized in Table~\ref{table:SysError}.
The largest systematic uncertainty comes from the extra $\pi^0$ veto.
The difference of peak width and
position between data and Monte Carlo produced the ``Peak shape'' systematic uncertainty.

\begin{table}[h]
\centering
\caption{Systematic uncertainty (all in percentage)}
\label{table:SysError}
\begin{tabular}{c|c|c|c}        \hline
Systematic &$D^0\to K^-\pi^+$ &$D^0\to K^-\pi^+ \pi^0$
&$D^0\to K^-\pi^+ \pi^+\pi^-$  \\ \hline
$\Delta E$ sideband   &0.14 &0.58 &0.57 \\
Background channel    &0.72 &0.98 &0.80 \\
Track simulation      &0.40 &0.0  &0.61 \\
Tag bias              &0.1  &0.1  &0.3  \\
Peak shape            &-0.90&2.44 &0.75 \\
Extra $\pi^0$ veto    &1.66 &1.66 &1.66 \\
$\eta$ veto           &-0.33&0.33 &0.75 \\\hline
$\pi^0$ efficiency    &  &  & \\\hline
All (without $\pi^0$ efficiency) &2.09 &3.18 &2.30 \\\hline
\end{tabular}
\end{table}

We have measured the braching fraction of $K_L^0 \pi^0$ in the tagged data sample,
and the branching fraction of $K_S^0 \pi^0$ in both tagged and untagged samples.
We use the two branching fractions of $K_S^0 \pi^0$ to get
the correction factor,$1-2r_fcos\delta+r_f^2$, and then apply the correction
factor, $1+2r_fcos\delta+r_f^2$, to the branching fraction of $D^0\to K^0_L \pi^0$
 to get the true branching fraction of $D^0\to K^0_L \pi^0$.

$r_f^2 \sim R_{WS}$ is taken from recent Belle results
\cite{Belle-1}, \cite{Belle-2}. By using the PDG value of ``y'', the branching fraction
for $D^0 \to K_S^0 \pi^0$ was corrected to be $1.202 \pm 0.016 \pm 0.039\%$. Then we
calculate $(1-2rcos\delta+r^2)$ for each tag mode separately. After correcting the branching
fractions for the three tags separately and combining the results together,
we get the branching fraction for $D^0 \to K_L^0 \pi^0$: $0.940 \pm 0.046 \pm 0.032\%$.  Note
that the $\pi^0$ systematic uncertainty is not included.

\section{Asymmetries Between $D \to \Ks\pi$ and $D \to \Kl\pi$}
To compare $D \to \Ks\pi$ and $D \to \Kl\pi$, we compute the asymmetries
\begin{equation} \label{eq:asymDef}
R(D) \equiv \frac{ \BF(D \to \Ks \pi) - \BF(D \to \Kl \pi) }{ \BF(D \to \Ks \pi) + \BF(D \to \Kl \pi) }
\end{equation}
The error propagation in the asymmetry is complicated by correlations between the branching fractions.  For example, the $D \to \Kl \pi$ measurements include an input $D \to \Ks \pi$ branching fraction, so the two branching fractions are anti-correlated.  Also, the $D^0$ measurements both include the same $\pi^0$ systematic, so this systematic cancels.

The $D^+$ asymmetry is
\begin{displaymath}
R(D^+) = 0.030 \pm 0.023 \pm 0.025 \textrm{.}
\end{displaymath}
The uncertainties are dominated by the $D^+ \to \Ks\pi^+$ measurement, which used only a 56 pb$^{-1}$ subset of the 281 pb$^{-1}$ data set.  We expect an updated result soon, so the uncertainties on the asymmetry will improve significantly.

The $D^0$ asymmetry is
\begin{displaymath}
R(D^0) = 0.122 \pm 0.024 \pm 0.030 \textrm{.}
\end{displaymath}
This systematic uncertainty will also improve when the $D^0\bar{D}^0$ cross section measurement is updated with the full 281 pb$^{-1}$ data set.

\section{Interpretation}
The asymmetry measurements allow a measurement of the strong phase between $\Amp(\KpiDCS)$ and $\Amp(\KpiCF)$ under reasonable theoretical assumptions.

The three Cabibbo-favored $D \to K\pi$ decays are described by an isospin 1/2 amplitude $A_{1/2}$, an isospin 3/2 amplitude $A_{3/2}$, and their relative phase $\delta_I$.  The amplitudes are
\begin{eqnarray}
\Amp(\KpiCF) &=& \sqrt{\frac{2}{3}}A_{1/2} + \sqrt{\frac{1}{3}}A_{3/2}e^{-i\delta_I} \\
\Amp(\Kzbarpiz) &=& -\sqrt{\frac{1}{3}}A_{1/2} + \sqrt{\frac{2}{3}}A_{3/2}e^{-i\delta_I}  \\
\Amp(\Kzbarpip) &=& \sqrt{3} A_{3/2}e^{-i\delta_I}
\end{eqnarray}
Without loss of generality, we may take $A_{1/2}$ and $A_{3/2}$ to be real and positive.

The four doubly-Cabibbo-suppressed $D \to K\pi$ decays are described by one isospin 3/2 amplitude $B_{3/2}$, and two isospin 1/2 amplitudes $B_{1/2}$ and $C_{1/2}$.  It is reasonable to assume that $B_{3/2}$ and $A_{3/2}$ are relatively real, and that $B_{1/2}$, $C_{1/2}$, and $A_{1/2}$ are relatively real.  We can then write the amplitudes for the four doubly-suppressed decays
\begin{eqnarray}
\Amp(\KpiDCS) &=&  \sqrt{\frac{2}{3}}(B_{1/2}+C_{1/2}) + \sqrt{\frac{1}{3}}B_{3/2}e^{-i\delta_I} \\
\Amp(\Kzpiz) &=& -\sqrt{\frac{1}{3}}(B_{1/2}+C_{1/2}) + \sqrt{\frac{2}{3}}B_{3/2}e^{-i\delta_I} \\
\Amp(\Kpiz) &=& -\sqrt{\frac{1}{3}}(B_{1/2}-C_{1/2}) + \sqrt{\frac{2}{3}}B_{3/2}e^{-i\delta_I} \\
\Amp(\Kzpip) &=&  \sqrt{\frac{2}{3}}(B_{1/2}-C_{1/2}) + \sqrt{\frac{1}{3}}B_{3/2}e^{-i\delta_I}
\end{eqnarray}

In our notation, $A_{1/2}$, $A_{3/2}$, $B_{1/2}$, $C_{1/2}$, and $B_{3/2}$ are all real, and the phase in any amplitude comes from $\delta_I$.  We thus have six parameters to describe seven decays.  A complete fit is underway, but results are not yet available.  Instead, we illustrate what will be forthcoming by making some approximations.

The exact expressions for the asymmetries (defined in equation \ref{eq:asymDef}) are
\begin{eqnarray}
R(D^0) &=& \frac{ 2[A_{1/2}(B_{1/2}+C_{1/2}) + 2A_{3/2}B_{3/2} - \sqrt{2}(A_{1/2}B_{3/2} + A_{3/2}(B_{1/2}+C_{1/2}))\cos\delta_i] }{|A_{1/2} - \sqrt{2}A_{3/2}e^{-i\delta_I}|^2 + |(B_{1/2}+C_{1/2}) - \sqrt{2}B_{3/2}e^{-i\delta_I}|^2} \label{eq:RDzExact} \\
R(D^+) &=&\frac{ 2 A_{3/2} (B_{3/2} + \sqrt{2}(B_{1/2}-C_{1/2})\cos\delta_i) }{ 3A_{3/2}^2 + |\sqrt{\frac{2}{3}}(B_{1/2}-C_{1/2}) + \sqrt{\frac{1}{3}}B_{3/2}e^{-i\delta_I}|^2 } \label{eq:RDpExact}
\end{eqnarray}

From the three Cabibbo-favored decay rates, one readily shows that $A_{3/2} \approx (1/4) A_{1/2}$ and $\delta_I \approx 90^\circ$.  Furthermore, we assume that $|\Amp(\Kzpi)|^2$ is negligibly small compared to $|\Amp(\Kzbarpi)|^2$.  Thus, in the denominators of the expressions for $R(D)$, the second terms (with only $B$'s and $C$'s) are ignored.  We make these approximations in what follows.  The asymmetries simplify to
\begin{eqnarray}
R(D^0) &\approx& \frac{8}{9} \left[\frac{ 2(B_{1/2}+C_{1/2}) + B_{3/2} }{A_{1/2}}\right] \label{eq:RDzApprox} \\
R(D^+) &\approx& \frac{ 8 B_{3/2} }{ 3A_{1/2} } \label{eq:RDpApprox}
\end{eqnarray}

The ratio of doubly suppressed to allowed decays $r_{K\pi}^2 \equiv \frac{\BF(\KpiDCS)}{\BF(\KpiCF)}$ is given by
\begin{equation}
r_{K\pi}^2 = \frac{|(B_{1/2}+C_{1/2}) + \sqrt{\frac{1}{2}}B_{3/2}e^{-i\delta_I}|^2}{|A_{1/2} + \sqrt{\frac{1}{2}}A_{3/2}e^{-i\delta_I}|^2}
\end{equation}
and this simplifies to
\begin{equation}
r_{K\pi}^2 \approx \frac{32}{33} \left[ \left( \frac{B_{1/2}+C_{1/2}}{A_{1/2}}\right)^2 + \frac{1}{2}\left(\frac{B_{3/2}}{A_{1/2}} \right) ^2 \right] \label{eq:rKpiApprox}
\end{equation}

Combining the (approximate) expressions for $R(D^0)$, $R(D^+)$, and $r_{K\pi}^2$, we have
\begin{equation}
r_{K\pi}^2 \approx \frac{9}{88} [ 3 R(D^0)^2 - 2 R(D^0)R(D^+) + R(D^+)^2 ]
\end{equation}
Thus, $R(D^0)$ and $R(D^+)$ must lie on an ellipse, whose size is set by $r_{K\pi}^2$.

\begin{figure}[tbp]
\begin{center}
 \includegraphics[width=1.0\textwidth]{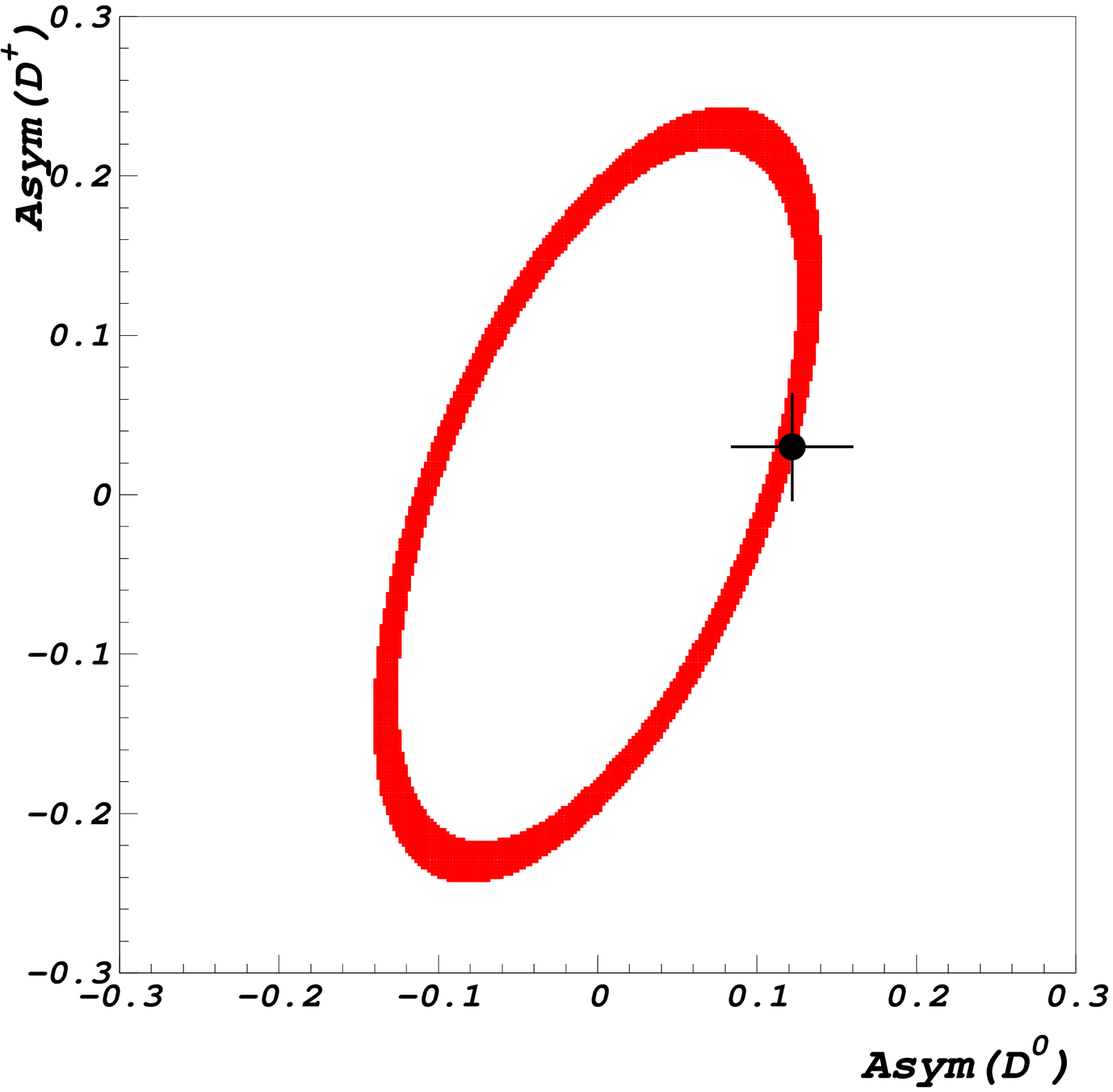}
\end{center}
\caption{Allowed values of $R(D^0)$ and $R(D^+)$.  The size of the ellipse is proportional to $r_{K\pi} = \sqrt{\BF(\KpiDCS)/\BF(\KpiCF)}$.  Our measurements are shown, with uncertainties.}
\label{fig:asymEllipse}
\end{figure}

Taking $r_{K\pi}^2 = 0.00363 \pm 0.00038$ from PDG 2004 branching fractions, we obtain the ellipse shown in Figure \ref{fig:asymEllipse}.  Our measured values for $R(D^0)$ and $R(D^+)$ are shown, and they lie on the ellipse, as they should.

\subsection{The $D^0 \to K^\pm \pi^\mp$ Strong Phase}
We can also determine the $D^0 \to K^\pm \pi^\mp$ strong phase, defined as the phase of the ratio of the two amplitudes
\begin{equation}
\frac{\Amp(\KpiDCS)}{\Amp(\KpiCF)} = r e^{i\delta_{\rm strong}}
\end{equation}
In particular,
\begin{equation}
r e^{i\delta_{\rm strong}} = \frac{ \sqrt{\frac{2}{3}}(B_{1/2}+C_{1/2}) + \sqrt{\frac{1}{3}}B_{3/2}e^{-i\delta_I} }{ \sqrt{\frac{2}{3}}A_{1/2} + \sqrt{\frac{1}{3}}A_{3/2}e^{-i\delta_I} }
\end{equation}
The simplified expression for $\delta_{\rm strong}$ is
\begin{equation}
\delta_{\rm strong} \approx \tan^{-1} \left[ \frac{1}{\sqrt{2}} \left( \frac{ \frac{1}{4}\frac{B_{1/2}+C_{1/2}}{A_{1/2}} - \frac{B_{3/2}}{A_{1/2}} }{ \frac{B_{1/2}+C_{1/2}}{A_{1/2}} + \frac{1}{8}\frac{B_{3/2}}{A_{1/2}} } \right) \right]
\end{equation}
Substituting $\frac{B_{1/2}+C_{1/2}}{A_{1/2}} \approx \frac{9}{16} R(D^0) - \frac{3}{16} R(D^+)$ and $\frac{B_{3/2}}{A_{1/2}} \approx \frac{3}{8}R(D^+)$,
\begin{equation}
\delta_{\rm strong} \approx \tan^{-1} \left[ \frac{1}{\sqrt{2}} \left( \frac{R(D^0)-3R(D^+)}{4R(D^0)-R(D^+)} \right) \right]
\end{equation}

Using the measured asymmetries as input, we find that the strong phase is consistent with zero:
\begin{displaymath}
\delta_{\rm strong} \approx (3 \pm 6 \pm 7)^\circ
\end{displaymath}
The quoted uncertainties do not include uncertainties due to the approximations $A_{3/2} \approx (1/4) A_{1/2}$ and $\delta_I \approx 90^\circ$.  These additional uncertainties would be at most a few degrees.

We have performed a preliminary study of fitting for the amplitude parameters with the $D \to K \pi$ rates as input.  This fit produces results consistent with the above approximations on $A_{3/2} / A_{1/2}$ and $\delta_I$.  We note that we can relax one assumption by allowing one additional phase to be non-zero -- for example, the phase of $C_{1/2}$ relative to $B_{1/2}$.  Studies of these more general fits are in progress.

\subsection{Constraint from $D^+ \to K^+ \pi^0$}
Consider the ratio of widths of the doubly-suppressed decays to charged kaon, $\Kpiz$ and $\KpiDCS$:
\begin{equation}
\rho \equiv \frac{\Gamma(\Kpiz)}{\Gamma(\KpiDCS)}
\end{equation}
Naively, one would expect this ratio to be 1/2.  Using the amplitudes given above,
\begin{equation} \label{eq:Rexact}
\rho = \frac{|-\sqrt{\frac{1}{3}}(B_{1/2}-C_{1/2}) + \sqrt{\frac{2}{3}}B_{3/2}e^{-i\delta_I}|^2}{|\sqrt{\frac{2}{3}}(B_{1/2}+C_{1/2}) + \sqrt{\frac{1}{3}}B_{3/2}e^{-i\delta_I}|^2}
\end{equation}
The naive result, $\rho = 1/2$, follows most simply from $|C_{1/2}/B_{1/2}| \ll 1$, $|B_{3/2}/B_{1/2}| \ll 1$ -- i.e., $C_{1/2} = B_{3/2} = 0$.   We could also have $|B_{1/2}/C_{1/2}| \ll 1$ instead.  A preliminary CLEO-c result, $\BF(D^+ \to K^+ \pi^0) = (2.25 \pm 0.36 \pm 0.15 \pm 0.07) \times 10^{-4}$ \cite{cleo-conf-06-10}, gives $\rho = 0.64 \pm 0.12$, consistent with the naive expectation.

Making the approximation $\delta_I \approx 90^\circ$, equation (\ref{eq:Rexact}) simplifies to
\begin{equation}
\rho \approx \frac{1}{2} \frac{(B_{1/2}-C_{1/2})^2 + 2 B_{3/2}^2}{(B_{1/2}+C_{1/2})^2 + \frac{1}{2} B_{3/2}^2}
\end{equation}
We can rewrite this as
\begin{equation} \label{eq:RapproxNoXY}
\rho \approx \frac{1}{2} \frac{ \left( 1 - 2 \frac{C_{1/2}}{B_{1/2}+C_{1/2}} \right) ^2 + 2 \left( \frac{B_{3/2}}{B_{1/2}+C_{1/2}} \right) ^2 }{1 + \frac{1}{2} \left( \frac{B_{3/2}}{B_{1/2}+C_{1/2}} \right) ^2}
\end{equation}
Call
\begin{eqnarray}
\frac{B_{3/2}}{B_{1/2}+C_{1/2}} &\equiv& \alpha \\
\frac{C_{1/2}}{B_{1/2}+C_{1/2}} &\equiv& \beta
\end{eqnarray}
Then equation (\ref{eq:RapproxNoXY}) becomes
\begin{equation} \label{eq:Rapprox}
\rho \approx \frac{1}{2} \frac{ ( 1 - 2 \beta ) ^2 + 2 \alpha^2 }{1 + \frac{1}{2} \alpha^2}
\end{equation}

Using the previously derived approximate expressions for $R(D^0)$ (\ref{eq:RDzApprox}) and $R(D^+)$ (\ref{eq:RDpApprox}), we find
\begin{equation}
\alpha = \frac{B_{3/2}}{B_{1/2}+C_{1/2}} = \frac{2R(D^+)}{3R(D^0) - R(D^+)}
\end{equation}

Taking our measured values $R(D^+) = 0.030 \pm 0.034$ and $R(D^0) = 0.122 \pm 0.038$, we have $\alpha = 0.18 \pm 0.24$, which leads to $\alpha^2 = 0.03^{+0.15}_{-0.03}$.

Treating $\alpha^2$ as a small number, we have
\begin{equation}
2\rho \approx (1-2\beta)^2 + \frac{3}{2} \alpha^2
\end{equation}
Using the values of $\rho$ and $\alpha$,
\begin{equation}
|1-2\beta| = 1.11^{+0.11}_{-0.16}
\end{equation}
So $\beta$ must be close to either 0 or 1.  If it is close to 0, $\beta = -0.06^{+0.08}_{-0.05}$.

If $\beta$ is close to zero, then $C_{1/2} \ll B_{1/2}$.  If it is near one, then $C_{1/2} \gg B_{1/2}$.  One needs theoretical arguments to decide between these two cases.

\section{Acknowledgements}
We gratefully acknowledge the effort of the CESR staff
in providing us with excellent luminosity and running conditions.
D.~Cronin-Hennessy and A.~Ryd thank the A.P.~Sloan Foundation.
This work was supported by the National Science Foundation,
the U.S. Department of Energy, and
the Natural Sciences and Engineering Research Council of Canada.


\begin{thebibliography}{99}

\bibitem{cleo-conf-06-10}  S.A. Dytman {\it et al.} (CLEO Collaboration), CLEO-CONF-06-10, contributed to ICHEP06 [arXiv:hep-ex/0607075]. 
\bibitem{bigi}   I.I.Bigi, H.Yamamoto, Phys. Lett. B {\bf 349}, 363 (1995).
\bibitem{bib:Dhad} Q. He {\it et al.}, Phys. Rev. Lett. \textbf{95}, 121801 (2005). 
\bibitem{Belle-1}  K. Abe {\it et al.} (Belle Collaboration), BELLE-CONF-0254, contributed to ICHEP2002 [arXiv:hep-ex/0208051]. 
\bibitem{Belle-2}  X. C. Tian {\it et al.} (Belle Collaboration), Phys. Rev. Lett. {\bf 95}, 231801 (2005). 

\end{thebibliography}
\end{document}